\documentclass[showpacs,preprintnumbers,prb]{revtex4}
\usepackage{graphicx}
\usepackage{dcolumn}
\usepackage{bm}

\input{tcilatex}

\begin{document}

\title{The Bean-Livingston barrier at a superconductor/magnet interface}
\author{Yu.~A.~Genenko}
\email{yugenen@tgm.tu-darmstadt.de}
\author{H.~Rauh}
\author{S.~V.~Yampolskii}
\altaffiliation[On leave from ]{Donetsk Institute for Physics and Technology, National Academy of Sciences
of Ukraine, 83114 Donetsk, Ukraine.}
\affiliation{Institut f\"{u}r Materialwissenschaft, Technische Universit\"{a}t Darmstadt,
D-64287 Darmstadt, Germany }
\date{\today}

\begin{abstract}
The Bean-Livingston barrier at the interface of type-II
superconductor/soft-magnet heterostructures is studied on the basis of the
classical London approach. This shows a characteristic dependence on the
geometry of the particular structure and its interface as well as on the
relative permeability of the involved magnetic constituent. The modification
of the barrier by the presence of the magnet can be significant, as
demonstrated for a cylindrical superconducting filament covered with a
coaxial magnetic sheath. Using typical values of the relative permeability,
the critical field of first penetration of magnetic flux is predicted to be
strongly enhanced, whereas the variation of the average critical current
density with the external field is strongly depressed, in accord with the
observations of recent experiments.
\end{abstract}

\maketitle

\section{Introduction}

Heterostructures on the macro or nano scales involving type-II
superconductor and ferromagnet elements show great potential for improving
superconductor properties such as critical currents and critical fields, and
therefore have been extensively studied both experimentally and
theoretically during the past few years~\cite%
{Schuller1,Schuller2,Moshch1,Moshch1a,Moshch2a,Moshch2,Moshch3,Peeters0,Pokrovsky1, Bulaevsky,Pokrovsky2,Pokrovsky2a,Johansen,Peeters,Peeters2,Sonin,Erdin, Genenko1,Genenko2,Campbell,Campbell2,Dou1,Dou2,Dou3,Dou3a,Dou4,Jooss}%
. If hard magnets are used, the interaction of the magnetic vortices of the
superconductor with the magnetic moments of the ferromagnet may lead to an
enhancement of the pinning of the vortices~\cite%
{Moshch1,Moshch1a,Moshch2a,Moshch2} or to an increase of the critical fields~%
\cite{Moshch3,Peeters2}. Soft magnets, on the other hand, aid to amend
superconductor performance by shielding the transport current self-induced
magnetic field as well as the externally imposed magnetic field~\cite%
{Genenko1,Genenko2,Campbell,GRS1,GRS2}. Superconductors encompassed with
such materials exhibit enlarged critical currents through wide ranges of the
strength of an applied field, when in the critical state~\cite%
{Dou1,Dou2,Dou3,Dou3a,Dou4}, and even overcritical currents, when in the
Meissner state~\cite{Jooss}. The latter state persists until the shielding
and/or transport currents, which push magnetic vortices into the
superconductor bulk, overcome the Bean-Livingston barrier against entry of
magnetic flux~\cite{Bean}; an impediment created by the (positive) Gibbs
free energy of the vortices themselves. This suggests the surmise that, due
to the induced magnetization, the presence of a soft magnet may alter the
characteristics of nucleation of a vortex at the superconductor/magnet
interface as compared to nucleation at the surface of a superconductor
facing vacuum, the phenomenon analysed hitherto~\cite%
{Bean,Galajko,Petukhov,Koshelev,Samokhvalov,Genenko4}.

Any theoretical study of the Bean-Livingston barrier at the interface of a
type-II superconductor/soft-magnet heterostructure, discerned by the
observable critical field of first penetration of magnetic flux and the
observable average critical current density of loss-free transport of
electric charge interlinked with it, must resolve two cardinal points:

\noindent (a) the dependence of these observables on the geometry of the
particular structure and its interface;

\noindent (b) the effect of the relative permeability of the involved
magnetic constituent.

Here, we exemplify both traits for an infinite flat and, respectively,
finite curved geometry of a type-II superconductor next to a soft-magnet
environment, adopting the classical London approach.

\section{Theory}

The magnetic induction $B$\ in the superconductor region around a vortex
obeys the London equation~\cite{deGennes} 
\begin{equation}
\mathbf{B}+\lambda ^{2}\mathbf{\nabla }\times \left( \mathbf{\nabla }\times 
\mathbf{B}\right) =\mathbf{Q},  \tag{1}
\end{equation}

\noindent with the London penetration depth $\lambda $ and the source $%
\mathbf{Q}$ at position $\mathbf{r}$ given by 
\begin{equation}
\mathbf{Q(r)}=\Phi _{0}\dint\limits_{vc}d\mathbf{s}\,\delta (\mathbf{s}-%
\mathbf{r}),  \tag{2}
\end{equation}

\noindent where $\Phi _{0}$ denotes the quantum of magnetic flux and $\delta 
$ means the Dirac delta function, the integration extending along the vortex
core. The magnetic field $\mathbf{H}$ in the magnet region and the magnetic
induction $\mathbf{B}$ in the entire space satisfy the Maxwell equations 
\begin{equation}
\mathbf{\nabla }\times \mathbf{H}=0\text{\quad and\quad }\mathbf{\nabla }%
\cdot \mathbf{H}=0.  \tag{3}
\end{equation}

To simplify later analysis, we postulate the relationship $\mathbf{B}=\mu
\mu _{0}\mathbf{H}$ in the region confined to the magnet itself, assuming a
(field-independent) relative permeability of the magnet, $\mu $, apart from
the permeability of free space, $\mu _{0}$. Furthermore, to avoid
complications arising from the proximity effect, we invoke the presence of
an insulating layer at the superconductor/magnet interface, of thickness
much smaller than the London penetration depth (as observed, e.g., in MgB$%
_{2}$/Fe composites~\cite{Dou4}) regarding, for mathematical convenience,
this layer as infinitesimally thin. Boundary conditions then imply
continuity of the tangential component of the magnetic field as well as of
the normal component of the magnetic induction when the interface between
the superconductor (S) and the magnet (M) is traversed: 
\begin{equation}
B_{t,S}=\mu _{0}H_{t,M}\quad \text{and\quad }B_{n,S}=\mu \mu _{0}H_{n,M}. 
\tag{4}
\end{equation}

\subsection{Infinite flat geometry}

The configuration addressed first is thought to consist of a superconductor
extending across the infinite half-space $-\infty <x<0$ and a soft magnet
extending across the infinite half-space $0<x<\infty $, their interface
occupying the plane $x=0$, with an externally imposed, homogeneous field $%
\mathbf{B}_{0}$ pointing in the $y$-direction of a cartesian coordinate
system $x,y,z$, as shown in Fig.~1. For this usually discussed geometry of
the Bean-Livingston barrier~\cite{Bean}, the presence of the magnet does not
affect the entry of a straight magnetic vortex parallel to the
superconductor/magnet interface. Indeed, the vortex self-field here has a
tangential component only, which vanishes at the interface owing to the
vortex image field; the magnetization of the magnet thus is preserved,
leaving the vortex field the same as that of a vortex near the flat surface
of a semi-infinite superconductor facing vacuum. Nevertheless, the situation
could change in the more realistic case of fluctuation penetration of a
magnetic vortex loop (see Fig.~1), since the magnet then might experience
additional magnetization, with a corresponding interaction energy
contributing to the barrier at the superconductor/magnet interface.

\begin{figure}[tbp]
\includegraphics{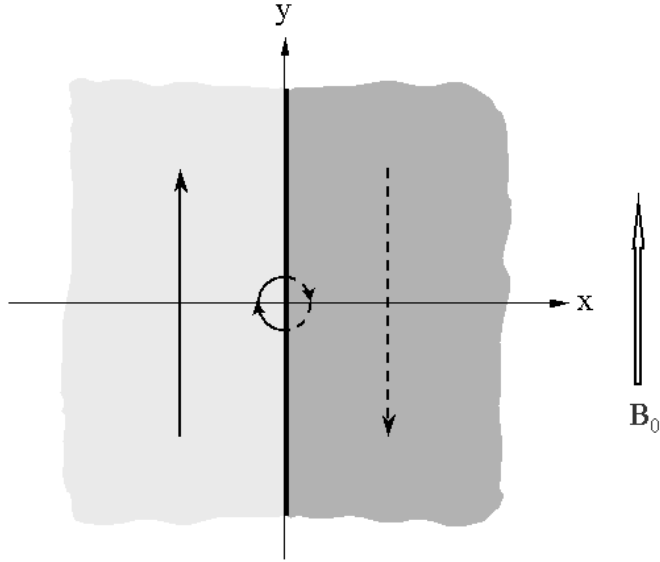}
\caption{Schematic view of the semi-infinite superconductor (light shading)
and the semi-infinite magnet (dark shading), their interface coinciding with
the plane $x=0$\ of a cartesian coordinate system $x,y,z$. The solid
vertical line indicates the insulating layer between the superconductor and
the magnet. A straight magnetic vortex parallel to the superconductor/magnet
interface (arrowed full line) and its image (arrowed dashed line) as well as
a magnetic vortex loop (arrowed full semicircle) and its image (arrowed
dashed semicircle), all situated in the plane $z=0$, are shown. The
direction of the external field $\mathbf{B}_{0}$ is marked.}
\end{figure}

The required solution of equations (1) - (4) for the magnetic induction $%
\mathbf{B}$ can be conveniently decomposed according to $\mathbf{B}=\mathbf{B%
}^{m}+\mathbf{b}$, where $\mathbf{B}^{m}$ is the Meissner field induced by
the external field in the absence of the magnetic vortex loop, 
\begin{equation}
B_{y}^{m}\left( x\right) =B_{0}\exp (x/\lambda ),  \tag{5}
\end{equation}

\noindent and $\mathbf{b}$ is the asymptotically vanishing field of the loop
itself. We represent the latter field and its source by two-dimensional
Fourier integrals of the kind 
\begin{equation}
\mathbf{f}\left( \mathbf{r}\right) =\frac{1}{\left( 2\pi \right) ^{2}}%
\int\limits_{-\infty }^{+\infty }dk_{y}\int\limits_{-\infty }^{+\infty
}dk_{z}\widetilde{\mathbf{f}}^{\left( k_{y},k_{z}\right) }\left( x\right)
\exp \left[ i\left( k_{y}y+k_{z}z\right) \right] .  \tag{6}
\end{equation}

\noindent For a vortex loop situated in the plane $z=0$, the Fourier
transforms of the cartesian components of the vortex field inside the
superconductor region are 
\begin{eqnarray}
\widetilde{b}_{x,y}^{\left( k_{y},k_{z}\right) }\left( x\right) &=&\left[
\left( \widetilde{b}_{x,y}^{\left( k_{y},k_{z}\right) }\left( 0\right) -%
\widetilde{P}_{x,y}^{\left( k_{y},k_{z}\right) }/2q\lambda ^{2}\right) %
\right] \exp \left( qx\right)  \nonumber \\
&&+\frac{1}{2q\lambda ^{2}}\int\limits_{0}^{a}dx^{\prime }\widetilde{Q}%
_{x,y}^{\left( k_{y},k_{z}\right) }\left( x^{\prime }\right) \exp \left(
-q\left\vert x+x^{\prime }\right\vert \right) ,  \TCItag{7} \\
\widetilde{b}_{z}^{\left( k_{y},k_{z}\right) }\left( x\right) &=&\widetilde{b%
}_{z}^{\left( k_{y},k_{z}\right) }\left( 0\right) \exp \left( qx\right) , 
\nonumber
\end{eqnarray}

\noindent where $q=\left( k^{2}+1/\lambda ^{2}\right) ^{1/2}$ with $k=\left(
k_{x}^{2}+k_{y}^{2}\right) ^{1/2}$; the boundary values herein read 
\begin{equation}
\widetilde{b}_{x}^{\left( k_{y},k_{z}\right) }\left( 0\right) =\frac{\mu 
\widetilde{P}_{x}^{\left( k_{y},k_{z}\right) }}{\left( k+\mu q\right)
\lambda ^{2}},\quad \widetilde{b}_{y,z}^{\left( k_{y},k_{z}\right) }\left(
0\right) =\frac{ik_{y,z}}{k}\frac{\widetilde{P}_{x}^{\left(
k_{y},k_{z}\right) }}{\left( k+\mu q\right) \lambda ^{2}},  \tag{8}
\end{equation}

\noindent with characteristic integrals of the type 
\begin{equation}
\widetilde{P}_{x,y}^{\left( k_{y},k_{z}\right)
}=\int\limits_{0}^{a}dx^{\prime }\widetilde{Q}_{x,y}^{\left(
k_{y},k_{z}\right) }\left( x^{\prime }\right) \exp \left( -qx^{\prime
}\right) ,  \tag{9}
\end{equation}

\noindent controlled by the shape of the loop and by the maximum distance
between the loop and the interface, $a$. The vortex field inside the magnet
region allows the representation $\mathbf{b}=-\mu \mu _{0}\nabla \psi $
through a scalar potential $\psi $, whose Fourier transform is given by 
\begin{equation}
\widetilde{\psi }^{\left( k_{y},k_{z}\right) }\left( x\right) =\frac{%
\widetilde{P}_{x}^{\left( k_{y},k_{z}\right) }}{k\left( k+\mu q\right)
\lambda ^{2}}\exp \left( -kx\right) .  \tag{10}
\end{equation}

We note that, although the magnetic moment of the vortex loop, with its only
component in the direction of the external field 
\begin{equation}
m_{y}=\int\limits_{0}^{a}dx^{\prime }\widetilde{Q}_{y}^{\left( 0,0\right)
}\left( x^{\prime }\right) \left[ 1-\exp \left( -x^{\prime }/\lambda \right) %
\right] ,  \tag{11}
\end{equation}

\noindent does not depend on $\mu $, the self-energy of the loop, being
determined by the field prevailing at the vortex core, could well be
sensitive to the magnetic environment. Yet, adopting a semicircular shape of
the loop with radius $a$, for which 
\begin{eqnarray}
\widetilde{Q}_{x}^{\left( k_{y},k_{z}\right) }\left( x\right) &=&2i\Phi
_{0}\sin \left( k_{y}\sqrt{a^{2}-x^{2}}\right) ,  \TCItag{12} \\
\Phi _{\mathbf{k}}^{y}\left( x\right) &=&-2\Phi _{0}\frac{x}{\sqrt{%
a^{2}-x^{2}}}\cos \left( k_{y}\sqrt{a^{2}-x^{2}}\right) ,  \nonumber
\end{eqnarray}

\noindent we find, using equations (7)-(10) in the region outside the vortex
core whose radial extent may be delineated by the coherence length of the
superconductor bulk, $\xi \ll \lambda $, the self-energy is represented by
the dominant term proportional to the length of the loop, 
\begin{equation}
F_{fi}\cong \left( \frac{\Phi _{0}^{2}}{4\pi \mu _{0}\lambda ^{2}}\right)
\pi a\ln \left( \frac{a}{\xi }\right) ,  \tag{13}
\end{equation}

\noindent apart from small corrections including the factor $1/\mu $ due to
the contribution of the magnet; a result which restates that for a
semi-infinite superconductor with a flat surface facing vacuum~\cite%
{Koshelev,Samokhvalov}. Intuitively, the decaying tendency of the omitted
corrections with increasing relative permeability here can be conceived in
the following way. The requirement of continuity of the normal component of
the magnetic induction across the superconductor/magnet interface on the one
hand, and the definition of the total magnetic flux through its quantization
in the superconductor on the other hand, ensure the strength of the magnetic
induction in the magnet region is typically $B\sim \Phi _{0}/\lambda ^{2}$,
whereas the strength of the magnetic field in the magnet region is typically 
$H\sim \Phi _{0}/\mu \mu _{0}\lambda ^{2}$. This yields a contribution to
the self-energy of the loop proportional to the product of both quantities,
which falls of as indicated above.

The Bean-Livingston barrier arises from a competition between attraction of
the vortex loop to the superconductor/magnet interface, accounted for by
equation (13), and repulsion due to the Lorentz force exerted~-- by the
Meissner current~-- on the loop. In the geometry of Fig.~1, this current
flows perpendicular to the plane $z=0$, and the work done by the external
field during growth of the loop from radius $a\cong \xi $ to radius $a\ll
\lambda $ is proportional to the area finally covered by the loop, 
\begin{equation}
\Delta W_{fi}\left( B_{0}\right) \cong \frac{1}{2}\Phi _{0}\pi
a^{2}j_{z}^{m}\left( B_{0}\right) ,  \tag{14}
\end{equation}

\noindent where 
\begin{equation}
j_{z}^{m}\left( B_{0}\right) =B_{0}/\mu _{0}\lambda  \tag{15}
\end{equation}

\noindent means the density of the Meissner current at the flat
superconductor/magnet interface. From equations (13) and (14), the Gibbs
free energy of the loop, i.e. the thermodynamic function of relevance here,
when $\xi \ll a\ll \lambda $, becomes 
\begin{equation}
G_{fi}\left( B_{0}\right) \cong \left( \frac{\Phi _{0}^{2}}{4\pi \mu
_{0}\lambda ^{2}}\right) \pi a\ln \left( \frac{a}{\xi }\right) -\frac{1}{2}%
\Phi _{0}\pi a^{2}j_{z}^{m}\left( B_{0}\right) ,  \tag{16}
\end{equation}

\noindent identifying the interface barrier against entry of the loop as a
function of the strength of the external field. Once, when $B_{0}$ is fixed,
the radius of the growing loop has reached its critical size $a_{c}$ defined
by the condition $\partial G_{fi}/\partial a=0$, further loop expansion
becomes irreversible and vortex entry proceeds. Depending on the quality of
the superconductor/magnet interface, this may happen at different values of
the critical loop radius throughout the range (and even beyond) where
equation (16) applies. Whilst for an ideal interface, with scale of
roughness $\sigma <\xi $, vortex entry occurs at a distance $a_{c}\cong \xi $%
, in the case of a real interface, with scale of roughness $\sigma \leq
\lambda $, vortex entry occurs for $a_{c}=\sigma $. Equation (16) in
conjunction with equation (15) thus yields for the critical field of first
penetration of magnetic flux across the flat superconductor/magnet interface 
\begin{equation}
B_{p}^{0}=\left( \frac{\Phi _{0}}{4\pi \lambda \sigma }\right) \ln \left( 
\frac{e\sigma }{\xi }\right) ,  \tag{17}
\end{equation}

\noindent a form which assumes values between $B_{c1}$ and $B_{c}$, the
lower and, respectively, thermodynamic critical field~\cite{deGennes}, when $%
\sigma $ varies between $\lambda $ and~$\xi $. Obviously, the interface
barrier against entry of the loop, and therefore the critical field of first
penetration of magnetic flux as well as the average critical current density
deriving from it, are insensitive to the magnetic environment, since in the
infinite configuration of Fig.~1, the external field remains totally
unshielded. However, in any finite superconductor/magnet heterostructure,
with the range of its magnetic constituent extending below the distance to
the sources of this field, a significant shielding effect may indeed occur.

\subsection{Finite curved geometry}

The configuration addressed next is thought to consist of a cylindrical
superconducting filament of radius~$R$, extended infinitely in the $z$%
-direction of a cartesian coordinate system $x,y,z$ and covered with a
coaxial magnetic sheath of thickness~$d$, the external field $\mathbf{B}_{0}$
again being aligned parallel to the $y$-direction, as depicted in Fig.~2. 
\begin{figure}[tbp]
\includegraphics{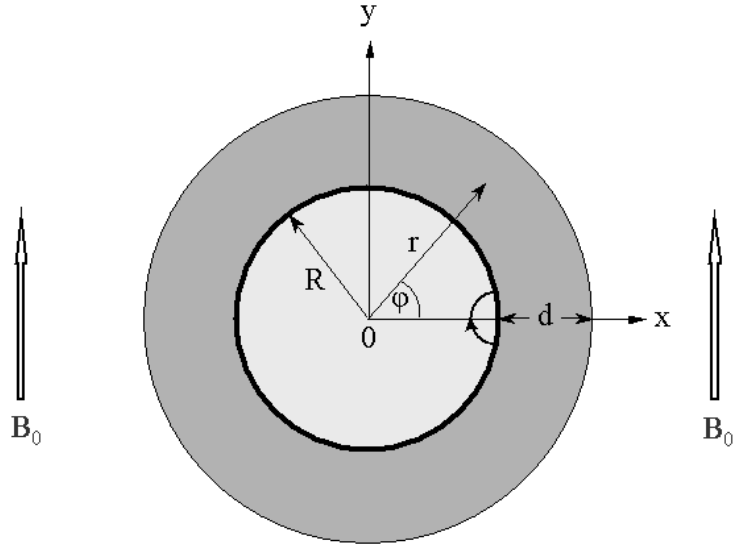}
\caption{Cross-sectional view of the superconducting filament of radius $R$
(light shading) and the coaxial magnetic sheath of thickness $d$ (dark
shading), their axes coinciding with the $z$-axis of a cartesian coordinate
system $x,y,z$. The solid ring indicates the insulating layer between the
superconducting filament and the magnetic sheath. A magnetic vortex loop
(arrowed semicircle, size not to scale!) situated in the plane $z=0$ and
nucleated at the filament/magnet interface is shown. The direction of the
external field $\mathbf{B}_{0}$ and the definition of cylindrical polar
coordinates $\left( r,\protect\varphi ,z\right) $ are marked.}
\end{figure}
\noindent Whereas the self-energy of an (almost) semicircular vortex loop of
radius $a\ll \lambda ,R$, located in the plane and nucleated at the
filament/magnet interface, duplicates the dominant term given by equation
(13), 
\begin{equation}
F_{ci}\cong \left( \frac{\Phi _{0}^{2}}{4\pi \mu _{0}\lambda ^{2}}\right)
\pi a\ln \left( \frac{a}{\xi }\right) ,  \tag{18}
\end{equation}

\noindent with corrections due to the effect of the magnetic sheath even
minor than those for the infinite configuration examined above~\cite%
{Genenko4,Genenko3}, the Meissner current, and hence the work done by the
external field during growth of the loop, as stated by equation (14), does
change markedly when considering the finite magnetic environment. We
therefore quote the Meissner solution of equation (1) expressed in
cylindrical polar coordinates $\left( r,\varphi ,z\right) $ adapted to the
filament. The radial and azimuthal components of the magnetic induction
inside the superconductor region are~\cite{VM} 
\begin{eqnarray}
B_{r}^{m}\left( r,\varphi \right)  &=&B_{0}A_{S}\left[ I_{0}\left( r/\lambda
\right) -I_{2}\left( r/\lambda \right) \right] \sin \varphi ,  \TCItag{19} \\
B_{\varphi }^{m}\left( r,\varphi \right)  &=&B_{0}A_{S}\left[ I_{0}\left(
r/\lambda \right) +I_{2}\left( r/\lambda \right) \right] \cos \varphi , 
\nonumber
\end{eqnarray}

\noindent with 
\begin{equation}
A_{S}=\frac{4\mu }{\left[ \left( \mu +1\right) ^{2}-\left( \mu -1\right)
^{2}R^{2}/\left( R+d\right) ^{2}\right] I_{0}\left( R/\lambda \right)
+\left( \mu -1\right) ^{2}\left[ 1-R^{2}/\left( R+d\right) ^{2}\right]
I_{2}\left( R/\lambda \right) },  \tag{20}
\end{equation}

\noindent where $I_{0}$\ and $I_{2}$ denote modified Bessel functions of the
first kind. Equations (19) in conjunction with Amp\`{e}re's law yield for
the density of the Meissner current flowing along the filament, 
\begin{equation}
j_{z}^{m}\left( r,\varphi \right) =2j_{z}^{m}\left( B_{0}\right)
A_{S}I_{1}\left( r/\lambda \right) \cos \varphi ;  \tag{21}
\end{equation}

\noindent an expression which adopts its maximum absolute value on the
circumference of the filament, $j_{\max }^{m}\left( B_{0}\right) $, at
angles $\varphi =0$ and~$\pi $ indicating the points of most probable
nucleation of the vortex loop. Accordingly, the work done by the external
field during growth of the loop from radius $a\cong \xi $ to radius $a\ll
\lambda $ is 
\begin{equation}
\Delta W_{ci}\left( B_{0}\right) \cong \frac{1}{2}\Phi _{0}\pi a^{2}j_{\max
}^{m}\left( B_{0}\right) ,  \tag{22}
\end{equation}

\noindent where, in the practically important limit $R\gg \lambda $, 
\begin{equation}
j_{\max }^{m}\left( B_{0}\right) \cong j_{z}^{m}\left( B_{0}\right) /\alpha 
\tag{23}
\end{equation}

\noindent represents the density of the Meissner current at the curved
filament/magnet interface, the parameter 
\begin{equation}
\alpha =\frac{1}{4}\left[ \mu +1-\left( \mu -1\right) \frac{R^{2}}{\left(
R+d\right) ^{2}}\right]  \tag{24}
\end{equation}

\noindent subsuming the combined shielding effect of the superconducting
filament and the magnetic environment. Thus, from equations (18) and (23),
the Gibbs free energy of the loop, say at the point $\varphi =0$, when $\xi
\ll a\ll \lambda $, becomes 
\begin{equation}
G_{ci}\left( B_{0}\right) \cong \left( \frac{\Phi _{0}^{2}}{4\pi \mu
_{0}\lambda ^{2}}\right) \pi a\ln \left( \frac{a}{\xi }\right) -\frac{1}{2}%
\Phi _{0}\pi a^{2}j_{\max }^{m}\left( B_{0}\right) ,  \tag{25}
\end{equation}

\noindent Evidently, the interface barrier against entry of the loop, taken
as a function of the strength of the external field, does prove sensitive to
the presence of the magnet owing to appreciable shielding of this field in
the finite configuration of Fig.~2. Again, referring to the condition $%
\partial G_{ci}/\partial a=0$ applied to equation (25) in conjunction with
equations (15), (17) and (23), we find for the critical field of first
penetration of magnetic flux across the curved filament/magnet interface, if 
$R\gg \lambda $, 
\begin{equation}
B_{p}\cong \alpha B_{p}^{0},  \tag{26}
\end{equation}

\noindent revealing a substantial enhancement by the factor~$\alpha $ as
compared to the respective field of first penetration of magnetic flux
across the flat superconductor/magnet interface for moderately high values
of the relative permeability already, brought about by the shielding effect.

The flow of a transport current, of magnitude~$i_{t}$, along the filament
means that the Meissner current density, given by equation (21), is to be
supplemented with the corresponding isotropic current density~\cite%
{Schafroth} 
\begin{equation}
j_{z,t}(r)=\frac{i_{t}}{2\pi R\lambda }\frac{I_{0}(r/\lambda )}{%
I_{1}(R/\lambda )},  \tag{27}
\end{equation}

\noindent and hence the maximum current density $j_{\max }^{m}\left(
B_{0}\right) $ entering equation (25) must be replaced by the maximum total
current density expressed, if $R\gg \lambda $, by 
\begin{equation}
j_{\text{tot}}\left( B_{0}\right) =j_{\max }^{m}\left( B_{0}\right) +\frac{%
i_{t}}{2\pi R\lambda }.  \tag{28}
\end{equation}

\noindent When $a_{c}$ is fixed, the condition $\partial G_{ci}/\partial a=0$
determines the average critical current density of loss-free transport along
the filament, $j_{c}=i_{t}/\pi R^{2}$, as well. Resorting to equations (23),
(25) and (28) in conjunction with equations (15) and (17), we get for $%
B_{0}<B_{p}$, if $R\gg \lambda $, 
\begin{equation}
j_{c}\left( B_{0}\right) \cong \frac{2\left[ B_{0}-B_{\max }^{m}\left(
B_{0}\right) \right] }{\mu _{0}R},  \tag{29}
\end{equation}

\noindent where 
\begin{equation}
B_{\max }^{m}\left( B_{0}\right) \cong B_{0}/\alpha ,  \tag{30}
\end{equation}

\noindent according to equations (19), reflects the maximum strength of the
magnetic induction at the curved filament/magnet interface. The average
critical current density thus is seen to fall off linearly with the strength
of the external field, at a rate determined by the parameter~$\alpha $, i.e.
by the shielding effect of both the superconducting filament and the
magnetic environment, revealing a strong reduction of the field dependence
for moderately high values of the relative permeability already, while the
zero-field value of the average critical current density is conserved.

We comment that, if the filament were absent, and hence shielding confined
to the magnetic sheath alone, the maximum strength of the magnetic induction
at the curved inner surface of the sheath would be decreased. By formally
letting $\lambda \rightarrow \infty $ for fixed $R$, equations (19) yield 
\begin{equation}
B_{\max }^{0}\left( B_{0}\right) \cong B_{0}/\beta ,  \tag{31}
\end{equation}

\noindent with the parameter 
\begin{equation}
\alpha =\frac{1}{4\mu }\left[ \left( \mu +1\right) ^{2}-\left( \mu -1\right)
^{2}\frac{R^{2}}{\left( R+d\right) ^{2}}\right] ,  \tag{32}
\end{equation}

\noindent duplicating an otherwise derived result~\cite{Batygin}. Since $%
\beta >\alpha $ holds for any value of the geometrical and material
characteristics involved, $B_{\max }^{0}\left( B_{0}\right) <B_{\max
}^{m}\left( B_{0}\right) $ ensues throughout, which confirms that the
predicted enhancement of the critical field, disclosed by equation (26),
like the concomitant attenuation of the external field, revealed by equation
(30), cannot be exclusively ascribed to the shielding effect of the magnetic
sheath, as argued in some previous attempts~\cite%
{Campbell2,DouAC1,DouAC2,DouAC3}.

To appraise the relevance of the above results, we take a MgB$_{2}$ filament
with radius $R=5.0\cdot 10^{-4}$~m covered by a Fe sheath of thickness $%
d=2.5\cdot 10^{-4}$~m and relative permeability $\mu =50$ (Refs.~\cite%
{Dou1,Dou2,Dou3}), noting the practically interesting temperature of $32$~K,
at which the London penetration depth and the coherence length adopt the
respective values $\lambda =1.8\cdot 10^{-7}$~m and $\xi =6.5\cdot 10^{-9}$%
~m (Ref.~\cite{Finnemore1}). If the scale of roughness~$\sigma $, and hence
the critical loop radius $a_{c}$, varies between the limits $\lambda $ and~$%
\xi $, the critical field of first penetration of magnetic flux across the
curved filament/magnet interface, $B_{p}$, given by equation (26), is found
to range between about $0.16$ and $1.02$~T (as compared to the range between
about $0.01$ and $0.07$~T when the magnetic sheath were absent), and the
average critical current density, $j_{c}\left( B_{0}\right) $, from equation
(29), turns out to vary between about $6.84\cdot 10^{7}$ and $4.42\cdot
10^{8}$~Am$^{-2}$ at zero external field, its rate of change with the field, 
$\partial j_{c}/\partial B_{0}$, amounting to about $-4.36\cdot 10^{8}$~Am$%
^{-2}$T$^{-1}$ (as opposed to about $-6.37\cdot 10^{9}$~Am$^{-2}$T$^{-1}$
when the magnetic sheath were absent). These estimates are in accord with
the observations of recent experiments~\cite{Dou1,Dou2,Dou3}. We add that,
considering the moderately large ratio of $\lambda /\xi $ around $28$, the
low critical temperature of about $40$~K and the just minor anisotropy of
polycrystalline MgB$_{2}$ (Ref.~\cite{Finnemore1}), thermally activated
penetration of magnetic flux across the barrier, considered before~\cite%
{Petukhov,Koshelev}, here is insignificant.

\section{Summary}

In conclusion, we have studied the Bean-Livingston barrier at the interface
of type-II superconductor/soft-magnet heterostructures and demonstrated a
characteristic dependence on the geometry of the particular structure and
its interface as well as on the relative permeability of the involved
magnetic constituent. Thus, for the flat interface between a semi-infinite
superconductor and a semi-infinite magnet, the external field remains
totally unshielded, leaving the barrier essentially the same as that at the
flat surface of a semi-infinite superconductor facing vacuum. However, in
any superconductor/magnet heterostructure, where substantial shielding of
the external field occurs, the modification of the barrier by the presence
of the magnet can be significant, as demonstrated for the example of a
cylindrical superconducting filament covered with a coaxial magnetic sheath.
In this finite geometry, with its curved superconductor/magnet interface,
using typical values of the relative permeability, we predict the critical
field of first penetration of magnetic flux is strongly enhanced and,
concomitantly, the variation of the average critical current density of
loss-free transport of electric charge with the external field is strongly
depressed; the zero-field critical current density value, however, is
retained, since the transport current self-induced magnetic field remains
unshielded in this geometry. Owing to the expulsion of magnetic flux out of
the filament, the attenuation of the external field, and hence the field
dependence of the average critical current density, cannot be ascribed to
the shielding effect of the magnetic environment alone.

\begin{acknowledgments}
Stimulating discussions with S.~X.~Dou, H.~C.~Freyhardt, A.~Gurevich,
J.~Horvat, H.~Jarzina, Ch.~Jooss, A.~V.~Pan, F.~M.~Peeters, M.~D.~Sumption
and V.~M.~Vinokur are gratefully acknowledged. This work was supported by
the VORTEX Programme of the European Science Foundation (ESF) and by a
research grant from the German Research Foundation (DFG).
\end{acknowledgments}

\bibliographystyle{plain}
\bibliography{apssamp}

\end{document}